\global\def\draftcontrol{0}

   \def\versionno{ kubogen -- draft   }
\catcode`\@=11

\expandafter\ifx\csname draftcontrol\endcsname\relax\global\def\draftcontrol{0}
\fi

{\count255=\time\divide\count255 by 60
\xdef\hourmin{\number\count255}
\multiply\count255 by-60\advance\count255 by\time
\xdef\hourmin{\hourmin:\ifnum\count255<10 0\fi\the\count255}}
\def\draftdate{\number\month/\number\day/\number\year\ \ \ \hourmin }

\newcommand\makepapertitle{\par
  \begingroup
    \renewcommand\thefootnote{\@fnsymbol\c@footnote}%
    \def\@makefnmark{\rlap{\@textsuperscript{\normalfont\@thefnmark}}}%
    \long\def\@makefntext##1{\parindent 1em\noindent
            \hb@xt@1.8em{%
                \hss\@textsuperscript{\normalfont\@thefnmark}}##1}%
     \newpage
     \global\@topnum\z@   % Prevents figures from going at top of page.
     \@makepapertitle
     \thispagestyle{empty}\@thanks
  \endgroup
  \setcounter{footnote}{0}%
  \global\let\thanks\relax
  \global\let\makepapertitle\relax
  \global\let\@makepapertitle\relax
  \global\let\@thanks\@empty
  \global\let\@author\@empty
  \global\let\@date\@empty
  \global\let\@title\@empty
  \global\let\title\relax
  \global\let\author\relax
  \global\let\date\relax
  \global\let\and\relax
  \def\version{\let\version\@version\@gobble}
}
\def\@makepapertitle{%
  \newpage
   \ifnum\draftcontrol=1 {}
   \version\versionno
   \vskip 3em%
   \else
   \hfill\hbox to 3cm {\parbox{4cm}{\@pubnum}\hss}%
   \vskip 3em%
   \fi
   \begin{center}%
   \let \footnote \thanks
     {\LARGE {\@title}}%
     \vskip 1.5em%
     {\normalsize%\large
       \lineskip .5em%
       \begin{tabular}[t]{c}%
         \@author
       \end{tabular}\par}%
     \vskip 1.5em%
     {\@bstract}%
     \end{center}%
     \vskip 1.5em
     \@date%
   \par
}

\gdef\@pubnum{}
\def\pubnum#1{%
  \gdef\@pubnum{#1}}

\gdef\@bstract{}
\def\Abstract#1{%
  \gdef\@bstract{%
   \parbox{\textwidth-0pc}{%
   \centerline{\bf Abstract}\penalty1000%
\kern.2cm%
\noindent%\abstractfont \baselineskip=12pt
\renewcommand\baselinestretch{1.0}%
{#1}}}
}

\def\ps@paper{\let\@mkboth\@gobbletwo%
     \ifnum\draftcontrol=1
	\def\@oddfoot{\hbox to \textwidth{\tiny \versionno \hfil\tiny\draftdate}%
	\hskip -\textwidth \hbox to \textwidth{\hfil\rm\thepage\hfil}}%
     \else\def\@oddfoot{\hbox to \textwidth{\hfil\rm\thepage\hfil}}
     \fi
     \let\@evenfoot\@oddfoot
}
\def\body{\clearpage
          \pagestyle{paper}
	}
\def\@version#1{\ifnum\draftcontrol=1
\typeout{}\typeout{#1}\typeout{}
\vskip3mm\centerline{\hbox{\fbox{\normalsize{\tt DRAFT -- #1 -- }
                   {\draftdate}}}}\vskip3mm
\fi}
\let\version\@version
\long\def\eqlabel#1{\ifnum\draftcontrol=1
                    \tag@false  % there are some problems with multline without this
                    \tag*{(\theequation) \hbox to -0.2cm{\hspace{0cm}\small{#1}\hss}}
                    \refstepcounter{equation}
                    \edef\@currentlabel{\theequation}
                    \ltx@label{#1}          % use old LaTeX \label instead of new definition
                    \else
                    \label{#1}
                    \fi
                    }
\let\st@bibitem\@bibitem
\let\st@lbibitem\@lbibitem
\ifnum\draftcontrol=1
  \def\@bibitem#1{%
    \st@bibitem{#1}\a@@label{#1}\ignorespaces}
  \def\@lbibitem[#1]#2{%
    \st@lbibitem[#1]{#2}\a@@label{#2}\ignorespaces}
  \def\a@@label#1{%
    \gdef\a@lab{\smash{\normalfont\small#1}}
    \ifvmode
      \if@inlabel
        \global\setbox\@labels\hbox{%
          \llap{\a@lab\let\a@lab\relax
                \kern\@totalleftmargin\kern\marginparsep}%
          \box\@labels}%
      \fi
    \fi}
\fi
\documentclass[12pt,letterpaper]{article}
\usepackage{amsmath,amssymb,array,calc,rotating,epsfig,psfrag}
\usepackage[nosort]{cite}
\ifnum\draftcontrol=1
\fi
\renewcommand\baselinestretch{1.25}
\renewcommand\section{\@startsection {section}{1}{\z@}%
                                   {-3.5ex \@plus -1ex \@minus -.2ex}%
                                   {2.3ex \@plus.2ex}%
                                   {\normalfont\large\bfseries}}
\renewcommand\subsection{\@startsection{subsection}{2}{\z@}%
                                   {-3.25ex\@plus -1ex \@minus -.2ex}%
                                   {1.5ex \@plus .2ex}%
                                   {\normalfont\normalsize\bfseries}}
\renewcommand\subsubsection{\@startsection{subsubsection}{3}{\z@}%
                                   {-3.25ex\@plus -1ex \@minus -.2ex}%
                                   {1.5ex \@plus .2ex}%
                                   {\normalfont\normalsize\it}}
\renewcommand\paragraph{\@startsection{paragraph}{4}{\z@}%
                                   {-3.25ex\@plus -1ex \@minus -.2ex}%
                                   {1.5ex \@plus .2ex}%
                                   {\normalfont\normalsize\bf}}

\numberwithin{equation}{section}

\def\ie{{\it i.e.}}
\def\eg{{\it e.g.}}

\def\revise#1       {\raisebox{-0em}{\rule{3pt}{1em}}%
                     \marginpar{\raisebox{.5em}{\vrule width3pt\
                     \vrule width0pt height 0pt depth0.5em
                     \hbox to 0cm{\hspace{0cm}{%
                     \parbox[t]{4em}{\raggedright\footnotesize{#1}}}\hss}}}}

\def\cala         {{\cal A}}

\def\calb         {{\cal B}}

\def\calf         {{\cal F}}

\def\calh         {{\cal H}}

\def\call         {{\cal L}}
\def\calm         {{\cal M}}
\def\caln         {{\cal N}}
\def\calo         {{\cal O}}

\def\del          {\partial}

\def\Im           {{\rm Im\hskip0.1em}}

\def\sqr#1#2{{\vcenter{\vbox{\hrule height.#2pt
 \hbox{\vrule width.#2pt height#1pt \kern#1pt
 \vrule width.#2pt}\hrule height.#2pt}}}}

\newcommand{\ft}[2]{{\textstyle{\frac{#1}{#2}}}}

\def\w{\omega}
\def\om{\Omega}
\def\a{\alpha}
\def\tg{\tilde{g}}
\def\t{\triangle}
\def\p{\varphi}
\catcode`\@=12

\begin{document}

%%%
%%%%%% text starts here
%%%%%%%%%

\title{On universality of stress-energy tensor correlation functions in supergravity
}

\pubnum{%
hep-th/0408095}
\date{August 2004}

\author{
Alex Buchel\\[0.4cm]
\it Perimeter Institute for Theoretical Physics\\
\it Waterloo, Ontario N2J 2W9, Canada\\[0.2cm]
\it Department of Applied Mathematics\\
\it University of Western Ontario\\
\it London, Ontario N6A 5B7, Canada\\
}

\Abstract{
Using the Minkowski space AdS/CFT prescription we explicitly compute in
the low-energy limit the two-point correlation function of the boundary
stress-energy tensor in a large class of type IIB supergravity
backgrounds with a regular translationally invariant horizon. The
relevant set of supergravity backgrounds includes all geometries which
can be interpreted via gauge theory/string theory correspondence as
being holographically dual to finite temperature gauge theories in
Minkowski space-times. The  fluctuation-dissipation theorem relates this
correlation function computation to the previously established
universality of the shear viscosity from supergravity duals, and to
the universality of the low energy absorption cross-section for
minimally coupled massless scalars into a general spherically
symmetric black hole. It further generalizes the latter results for
the supergravity black brane geometries with non-spherical horizons.
}

%\enlargethispage{1.5cm}

\makepapertitle

\body

\version\versionno

\section{Introduction}
In the framework of gauge theory/string theory  correspondence \cite{m9711} 
the prescription for the computation of the Lorentzian-signature boundary gauge theory correlation 
functions was formulated in \cite{ss,hs}. This development enabled study of interesting non-equilibrium
processes (\eg\ diffusion and sound propagation) in strongly coupled thermal gauge theories 
\cite{ne1,ne2,ne3,ne4,ne5,bn2,bls}. In this paper we explicitly compute boundary stress-energy tensor retarded two-point correlation
function in a large class of type IIB supergravity backgrounds with regular translationally 
invariant horizon (``black branes''). We find that in the low-energy limit this correlation function has a universal 
dependence on the area of the horizon. The class of relevant supergravity geometries is generic enough to 
include  all geometries holographically dual to finite temperature gauge theories. For the latter 
subset, the fluctuation-dissipation theorem relates the universal properties of the boundary stress-energy tensor correlation 
functions to   the previously established universality of the shear viscosity in the effective hydrodynamic description 
of hot gauge theory plasma \cite{mp1,mp2}. Finally, for the black brane geometries that allow for an extension to asymptotically 
flat space-times\footnote{This is known to be the true for black brane geometries dual to finite temperature 
maximally supersymmetric gauge theories. We believe that this is true for all supergravity backgrounds holographically dual 
to gauge theories.} the universality of the correlation functions can be related \cite{c1,c2} to the 
universality of low energy absorption cross sections for black holes observed in \cite{das}. 
 
In the next section we introduce our conventions and describe type IIB supergravity 
backgrounds where boundary stress-energy tensor correlators exhibit universal properties.        
The computation of the correlation functions is delegated to section 3.
Appendix contains details of the derivation of the effective bulk action for the 
graviton fluctuations in supergravity backgrounds of section 2.

\section{Description of relevant supergravity backgrounds}
Consider static  type IIB supergravity backgrounds supported by various fluxes and/or 
axiodilaton with a regular horizon\footnote{By a 'regular horizon' we mean singularity-free horizon of finite area.}.
We assume that Einstein frame ten-dimensional geometry 
is a direct warped product of the time direction, $p>2$ -dimensional Euclidean space $R^p$, and an
arbitrary $q=9-p$ -dimensional (noncompact) 'transverse space' $\calm_q$ 
\begin{equation}
ds_{10}^2\equiv\hat{g}_{MN}d\xi^Md\xi^N=-\om_1^2(y)\ dt^2+\om_2^2(y)\left(dx^\a dx^\a+\tg(y)_{mn}dy^mdy^n \right)\,,
\eqlabel{metric}
\end{equation} 
where $\a=1\dots p$ labels Euclidean directions and $\tg_{mn}$ is the metric on $\calm_q$.
The transverse manifold $\calm_q$ is assumed to be singularity-free\footnote{$\calm_q$ does not 
have to be a Calabi-Yau (CY) manifold. In fact, in all explicit examples of gauge/string
duality at finite temperature $\calm_q$ is not a CY space, as the supersymmetry of the dual gauge 
theory is broken.  
} 
and to have only one $(q-1)$-dimensional  boundary component $\del\calm_q$.
Also, we assume that curvature invariants of the full ten-dimensional metric are small in Planck units so 
that the supergravity approximation is valid.      
It turns out, the key property of the background geometry pertinent 
to the universality of the correlation functions is the following relation
between components of  the Ricci tensor of \eqref{metric}
\begin{equation}
R_t^t-R_{\a}^\a=0\,,
\eqlabel{rel}
\end{equation}  
where there is no summation over $\a$. As emphasized in \cite{mp2}, \eqref{rel} 
is automatically satisfied for all supergravity geometries holographically dual 
to strongly coupled finite temperature gauge theories in $R^{1,p}$ Minkowski space-time.
The argument goes as follows \cite{mp2}. First observe that for extremal 
(zero temperature) backgrounds the Poincar\'e symmetry of the background geometry ensures that 
the longitudinal components of the stress tensor can only have the form $T_{\mu\nu}\sim
g_{\mu\nu}(\dots)$. Next note that, while turning on nonextremality involves modifications to 
the metric as well as to the profile of matter fields over $\calm_q$, this has no effect on the 
structure of $T_{\mu\nu}$. Given explicit expression for the 
type IIB supergravity matter stress tensor \cite{s83}, the latter is a trivial consequence of the fact 
that even off the extremality 
the axiodilaton and fluxes vary only over $\calm_q$, and 3-form fluxes are transverse to $R^{1,p}$
at the extremality.  Thus, $T_t^t-T_\a^ \a=0$ for both extremal and nonextremal backgrounds.  
The Einstein equation then gives \eqref{rel}. 

There are interesting non-extremal supergravity geometries which do not honor \eqref{rel}. One example is 
supergravity dual to finite temperature $\caln=4$ $SU(N)$ supersymmetric Yang-Mills theory with a 
nonzero chemical potential for a $U(1)\subset SO(6)_R$ R-charge.   
Kaluza-Klein reduction on the five-sphere of the corresponding supergravity solution is  
the STU-model \cite{cv1} of the five-dimensional gauged supergravity. Here (see eqs.~(24), (25)  of \cite{cv1})
\begin{equation}
R_t^t-R_\a ^\a=\frac 12 F^2\,,
\eqlabel{cvf1}
\end{equation}
where $F^2=F_{MN}F^{MN}$ is a square of the field strengths of the five-dimensional $U(1)$ gauge potential 
corresponding to an R-charge chemical potential. As explained in \cite{cv2}, $F^2\ne 0$, leading to violation 
of \eqref{rel}. The reason for such a violation is quite simple from the ten-dimensional perspective.
Indeed, the 10d uplift of the STU models represents spinning extremal D3 branes \cite{cv2}, which metric involves 
a cross-term of the form $dt d\phi$ ($\phi$ is one of the $S^5$ coordinates), sourced by the 5-form stress tensor 
$T_{\mu\nu}\not\sim g_{\mu\nu}(\cdots)$.     
As a result, we expect that in this geometry the boundary stress-energy tensor correlation 
function will not have a universal form derived in section 3. This in turn implies the deviation from 
the universal result for the shear viscosity  \cite{mp2}. Needless to say, it will be interesting to 
explicitly compute this deviation and verify the Kovtun, Son and Starinets (KSS) shear viscosity bound \cite{mp1}.

In what follows we restrict our attention to the geometries satisfying constraint \eqref{rel}.
Also we take $p=3$ ($q=6$). Other cases ($p\ne 3$) can be studied along the same lines, and lead to 
identical conclusions. We take the following ansatz for type IIB supergravity matter fields, 
which is compatible (through Einstein equations) with \eqref{rel}. Both the axiodilaton $\tau=\tau(y)\equiv C_{(0)}+i e^{-\phi}$
and the 3-form fluxes $G_3=G_3(y)\equiv F_3-\tau H_3$ vary\footnote{Here and below 
we use $y$ to denote collection of coordinates $\{y^m\}$ on $\calm_q$.} only over $\calm_6$. Additionally $G_3$ 
has nonvanishing components only along $\calm_6$.  For the 5-form $\calf_5$ we assume 
\begin{equation}
\calf_5=(1+\star)[d\w\wedge dt\wedge dx^1\wedge dx^2\wedge dx^3]\,,
\eqlabel{5form}
\end{equation}
with $\w=\w(y)$.  Explicit computation of the Ricci tensor components of \eqref{metric} yields \cite{mp2}
\begin{equation}
\begin{split}
R_t^t=&\om_2^{-2}\om_1^{-1} \nabla^2\om_1+7\om_2^{-3}\om_1^{-1}\nabla\om_1\nabla\om_2\,,\\
R_\a^{\a}=&\om_2^{-3}\nabla^2\om_2+6\om_2^{-4}\left(\nabla\om_2\right)^2+\om_2^{-3}\om_1^{-1}\nabla\om_1\nabla\om_2\,,\\
\end{split}
\eqlabel{ri}
\end{equation}   
where $\nabla$ is with respect to $\tg_{mn}$. It will be convenient to introduce $\t(y)$ as
\begin{equation}
\om_1(y)=\om_2(y)\t(y)\,.
\eqlabel{tdef}
\end{equation} 
Given \eqref{ri} we find from \eqref{rel} \cite{mp2}
\begin{equation}
0=\nabla^2\t+8\om_2^{-1}\nabla\om_2\nabla\t=\om_2^{-8}\nabla\left(\om_2^8\nabla\t\right)\,.
\eqlabel{rel1}
\end{equation}

We assume that supergravity geometry \eqref{metric} is that of the black brane with a regular 
Schwarzschild horizon. Horizon of \eqref{metric} is an eight-dimensional submanifold with a 
direct product structure $R^3\times \calh_5$, where $\calh_5$ is a co-dimension one submanifold of 
$\calm_6$ determined by 
\begin{equation}
\t\bigg|_{\calh_5}=0\,.
\eqlabel{hordef}
\end{equation}
Regularity of the horizon requires that 
\begin{equation}
\om_2\bigg|_{\calh_5}\ne 0\,,\qquad \det(h_{mn})\equiv h\bigg|_{\calh_5}\ne 0\,,
\end{equation}
where $h_{mn}$ is the induced metric on $\calh_5$.
Notice that we do not require $\calh_5$ to be 'spherical', \ie, depend only on a 'radial' coordinate 
of $\calm_6$. As we approach the boundary $\del\calm_6$ of $\calm_6$ we expect the restoration of the 
four-dimensional Poincar\'e symmetry. This leads to 
\begin{equation}
\t\bigg|_{\del\calm_6}=1\,.
\eqlabel{tinf}
\end{equation}  
The {\it zero law} of black hole (brane) thermodynamics requires \cite{peet} that the 
surface gravity (or equivalently the temperature) is constant over the horizon
for a stationary black hole. This can be easily understood as a requirement for the 
absence of conical singularity in the analytically continued Euclidean geometry 
$t\to t_E=it$. Let $n^n$ be a unit normal vector to $\calh_5$ in $\calm_6$, \ie,
\begin{equation}
n^n n^m\tg_{nm}\bigg|_{\calh_5}=1\,,\qquad n^n v^m\tg_{nm}\bigg|_{\calh_5}=0\,, 
\eqlabel{normal}
\end{equation} 
for any vector $\{v^m\}\in \calh_5$. 
From the definition of $\calh_5$ as a horizon \eqref{hordef}
\begin{equation}
\nabla\t-\left(n\nabla\t\right)n\bigg|_{\calh_5}=0\,.
\eqlabel{rel2}
\end{equation}
Moreover, the zero law of black brane thermodynamics implies that 
\begin{equation}
n\nabla\t\bigg|_{\calh_5}={\rm const}=2\pi T\,,
\eqlabel{rel3}
\end{equation}
where $T$ is the Hawking's temperature of the black brane. 

We conclude this section with two observations useful for the evaluation of the stress-energy tensor
correlation function. First, notice that from \eqref{rel2}, \eqref{rel3}
\begin{equation}
\left(\nabla\t\right)^2\bigg|_{\calh_5}={\rm const}=\left(2\pi T\right)^2\,.
\eqlabel{ob1}
\end{equation}
Second, the area $\cala_8$ of the black brane horizon is
\begin{equation}
\cala_8=V_3\int_{\calh_5}d^5\xi\sqrt{h}\ \om_2^8\,,
\eqlabel{area}
\end{equation} 
where $V_3$ is  (a divergent) area of $R^3$.

\section{Boundary stress-energy tensor two-point correlation functions}
In this section,  using prescription \cite{ss}, we compute retarded Green's  
function of the boundary stress-energy tensor $T_{\mu\nu}(t,x^\a)$ ($\mu=\{t,x^\a\}$)
at zero spatial momentum, and in the low-energy 
limit $\w\to 0$, 
\begin{equation}
G^{R}_{12,12}(\w,0)=-i\int dt d^3x\ e^{i\w t}\theta(t)\langle[T_{12}(t,x^\a),T_{12}(0,0)]\rangle\,.
\eqlabel{defgr}
\end{equation}
We find 
\begin{equation}
G^{R}_{12,12}(\w,0)=-\frac{i\w s}{4\pi}\left(1+\calo\left(\frac{\w}{T}\right)\right)\,,
\eqlabel{result}
\end{equation}
where 
\begin{equation}
s=\frac{\cala_8}{4V_3 G_N}
\eqlabel{entdef}
\end{equation}
is the Bekenstein-Hawking entropy density of the black brane. $G_N=\frac{k_{10}^2}{8\pi}$ is a ten-dimensional Newton constant.
Given \eqref{result} and the Kubo relation 
\begin{equation}
\eta=\lim_{\w\to 0}\frac{1}{2\w i}\left[G^A_{12,12}(\w,0)-G^R_{12,12}(\w,0)\right]\,,
\eqlabel{kubo}
\end{equation}
where $G^A(\w,0)=\left(G^R(\w,0)\right)^\star$ is advanced Green's function,
we can reproduce the universality of shear viscosity $\eta$ of strongly 
coupled gauge theories from supergravity  \cite{mp2}
\begin{equation}
\frac{\eta}{s}=\frac{1}{4\pi}\,.
\eqlabel{sv}
\end{equation}

We begin computation of \eqref{defgr}
recalling that the coupling between the boundary value of the graviton 
and the stress-energy tensor of a gauge theory is given by $\delta g_2^1T_1^2/2$. According to the gauge/gravity prescription,
in order to compute the retarded thermal two-point function \eqref{defgr} we should add a small bulk perturbation 
$\delta g_{12}(t,y)$ to the metric \eqref{metric}, and compute the on-shell action as a functional of its  
boundary value $\delta g_{12}^b(t)$. Simple symmetry arguments \cite{ne2} show that for a perturbation of this type and 
metric of the form \eqref{metric} all other components of a generic perturbation $\delta g_{\mu\nu}$ 
can be consistently set to zero. It will be convenient to introduce a field $\p=\p(t,y)$,
\begin{equation}
\p=\frac 12 g^{\a\a}\ \delta g_{12}=\frac 12 \om_2^{-2}\ \delta g_{12}\,.
\eqlabel{defp}
\end{equation}
Following \cite{ss,hs}, retarded correlation function $G^R_{12,12}(\w,0)$ can be extracted from the 
(quadratic) boundary effective action $S_{boundary}$ for the metric fluctuations $\p^b$,
\begin{equation}
\p^b(\w)=\int \frac{d^4k}{(2\pi)^4} e^{-i\w t}\
\p(t,y)\bigg|_{\del\calm_6}\,,
\eqlabel{pb}
\end{equation} 
given by 
\begin{equation}
S_{boundary}[\p^b]=\int \frac{d^4k}{(2\pi)^4}\ \p^{b}(-\w)\ \calf(\w,y)\ \p^{b}(\w)\bigg|^{\del\calm_6}_{\calh_5}\,,
\eqlabel{sssb}
\end{equation}
as 
\begin{equation}
G^R_{12,12}(\w,0)=\lim_{\del\calm_6^r\to\del\calm_6}\ 2\ \calf^r(\w,y)\,. 
\eqlabel{Gr}
\end{equation}
The boundary metric functional is defined as
\begin{equation}
S_{boundary}[\p^b]=\lim_{\del\calm_6^r\to\del\calm_6}\biggl(
 S^r_{bulk}[\p]+S_{GH}[\p]+S^{counter}[\p]\biggr)\,,
\eqlabel{bounfu}
\end{equation}
where $S^r_{bulk}$ is the bulk 
Minkowski-space type IIB supergravity  action on a cut-off space: $\calm_6$  in \eqref{metric}
is regularized by a compact manifold $\calm_6^r$ with a boundary $\del\calm_6^r$. 
Also, $S_{GH}$ is the standard Gibbons-Hawking term over the regularized boundary $\del\calm_6^r$. 
The regularized bulk action $S^r_{bulk}$ is evaluated on-shell 
for the bulk metric fluctuations $\p(t,y)$ subject to the following boundary conditions:
\begin{equation}
\begin{split}
&(a):\ \lim_{\del\calm_6^r\to\del\calm} \p(t,y)=\p^b(t)\,,\\
&(b):\ \p(t,y)\ {\rm  is\ an\ incoming\ wave\ at\ the\ horizon}\ \calh_5\,.  
\end{split}
\eqlabel{bc}
\end{equation}  
The purpose of the boundary counterterm $S^{counter}$  is to remove 
divergent (as $\del\calm_6^r\to\del\calm_6$) and $\w$-independent contributions from the
kernel $\calf$ of \eqref{sssb}. 

Effective bulk action for $\p(t,y)$ in supergravity backgrounds specified in previous section
(derivation details are given in Appendix) takes the following form
\begin{equation}
\begin{split}
&S_{bulk}[\p]=\frac{1}{2k_{10}^2}\int d^{10}\xi\ \om_1\om_2^9\sqrt{\tg}\biggl[
\om_1^{-2}\left\{\ft 12 \left(\del_t \p\right)^2-\del^2_t\left(\p^2\right)\right\}\\
&+\om_2^{-2}\left\{-\ft 12 \left(\nabla\p\right)^2+\nabla^2\left(\p^2\right)+\nabla(\ln\om_1)
\nabla\left(\p^2\right)+8\nabla(\ln\om_2)
\nabla\left(\p^2\right)\right\}\\
&+\p^2\om_1^{-1}\om_2^{-9}\nabla\left(\om_2^7\nabla\om_1\right)
\biggr]\,,
\end{split}
\eqlabel{ac1}
\end{equation}
or equivalently 
\begin{equation}
\begin{split}
S_{bulk}[\p]\equiv&\int d^{10}\xi\ \call_{10}
=\frac{1}{2k_{10}^2}\int d^{10}\xi\ \biggl[\\
&\om_1\om_2^9\sqrt{\tg}
\biggl\{\frac{\om_1^{-2}}{2} \left(\del_t \p\right)^2-\frac{\om_2^{-2}}{2} \left(\nabla\p\right)^2\biggr\}\\
&+\biggl\{-\del_t\left(\sqrt{\tg}\om_1^{-1}\om_2^9\ \del_t(\p^2)\right)
+\nabla\left(\sqrt{\tg}\om_1\om_2^7\ \nabla(\p^2)\right)+\nabla\left(\sqrt{\tg}\p^2\om_1\om_2^6\ \nabla(\om_2)\right)
\biggr\}\\
&+\p^2\sqrt{\tg}\biggl\{\nabla\left(\om_2^7\nabla\om_1\right)-\nabla\left(
\om_1\om_2^6\nabla\om_2\right)\biggr\}
\biggr]\,.
\end{split}
\eqlabel{ac2}
\end{equation}
The second line in \eqref{ac2} is the effective action for minimally coupled scalar in geometry \eqref{metric},
the third line is a total derivative. Finally, given \eqref{rel1}, the last line in \eqref{ac2} identically 
vanishes. Thus, bulk  equation of motion for $\p$  is that of a minimally coupled scalar in \eqref{metric}. 
Decomposing $\p$ as 
\begin{equation}
\p(t,y)=e^{-i\w t}\p_\w(y)\,,
\end{equation} 
we find
\begin{equation}
\nabla\left(\om_1\om_2^7\nabla\p_\w\right)+\w^2\om_1^{-1}\om^9_2\p_\w=0\,.
\eqlabel{eomp}
\end{equation}
Similar to \cite{bn2}, a low-frequency solution of \eqref{eomp} which is an incoming wave at the horizon, 
and which near the boundary satisfies 
\begin{equation}
\lim_{\del\calm_6^r\to\del\calm_6}\ \p_\w(y)=1\,,
\eqlabel{blim}
\end{equation} 
can be written as 
\begin{equation}
\p_\w(y)=\t^{-i\w Q}\ \biggl(F_0+i\w Q\ F_{\w}+\calo(\w^2)\biggr)\,,
\eqlabel{taylorsol}
\end{equation} 
where $\t$ is defined as in \eqref{tdef} and $F_0=F_0(y)$, $F_\w=F_\w(y)$.
Exponent $Q>0$ determines a leading  singularity of $\p_w$ at the horizon. We find 
\begin{equation}
Q=\frac{1}{\sqrt{(\nabla\t)^2}}\bigg|_{\calh_5}=\frac{1}{2\pi T}\,,
\eqlabel{defq}
\end{equation}
where we used \eqref{ob1}. Notice that the fact that $Q$ is constant is related to the zero law 
of the black brane thermodynamics!
Smooth at the horizon functions   $\{F_0,\ F_{\w}\}$ satisfy  following  
partial differential equations 
:
\begin{equation}
\begin{split}
0=&\nabla\left(\t\om_2^8\nabla F_0\right)\,,\\
0=&\nabla\left(\t\om_2^8\nabla F_w\right)-2\om_2^8\nabla\t\nabla F_0\,,
\end{split}
\eqlabel{difff}
\end{equation}
with the general solution (recall \eqref{rel1}) 
\begin{equation}
\begin{split}
&F_0=c_0+c_1\ \ln\t\,,\\
&F_\w=c_1\ (\ln\t)^2+c_2\ \ln \t +c_3\,,
\end{split}
\eqlabel{difff1}
\end{equation}
where  $c_i$ are integration constants.
The only  solution  \eqref{difff1} nonsingular at the horizon which also  
 satisfies \eqref{blim} is  
\begin{equation}
F_0(y)= 1\,,\qquad F_{\w}(y)= 0\,.
\eqlabel{solf}
\end{equation}
Thus, 
\begin{equation}
\p(t,y)=e^{-i\w t}\ \t^{-i\w Q}\ \left(1+\calo(\w^2)\right)\,.
\eqlabel{solff}
\end{equation}

Once the bulk fluctuations are on-shell (\ie,\ satisfy  equations of motion)
the bulk gravitational Lagrangian becomes a total derivative. From \eqref{ac2} 
we find (without dropping any terms)
\begin{equation}
2k_{10}^2\ \call_{10}=\del_t J^t+\nabla J^y\equiv\del_t J^t+\nabla_m \left(\tg^{mn}J^y_n\right)\,,
\eqlabel{tder}
\end{equation}
where 
\begin{equation}
\begin{split}
J^t=&-\frac 32 \sqrt{\tg}\om_1^{-1}\om_2^9\ \p\del_t\p\,, \\
J^y=&\frac 32 \sqrt{\tg}\om_1\om_2^7\ \p\nabla\p+\sqrt{\tg}\om_1\om_2^6\nabla\om_2\ \p^2\,.
\end{split}
\eqlabel{jr}
\end{equation}
Additionally, the Gibbons-Hawking term provides an extra contribution so that 
\begin{equation}
J^y\to J^y-2 \sqrt{\tg}\om_1\om_2^7\ \p\nabla\p\,.
\eqlabel{ghshift}
\end{equation}

We are now ready to extract the kernel $\calf$ of \eqref{sssb}.   
The regularized boundary effective action for $\p$ is
\begin{equation}
\begin{split}
S_{boundary}[\p]^r=&S^r_{bulk}[\p]+S_{GH}[\p]+S^{counter}[\p]\\
=&\frac{1}{2k_{10}^2}\int dtd^3x\ \int_{\calm_6^r}d^6y\ \left(\del_tJ^t+\nabla J^y\right)+
\int dtd^3x\ \int_{\del\calm_6^r}d^5\xi\ \call^{counter}[\p]\\
=&\frac{1}{2k_{10}^2}\int dtd^3x\ \int_{\del\calm_6^r}d^5\xi\ \sqrt{h}\biggl(
-\frac 12 \om_1\om_2^7\ N^n \del_n\p\ \p\\
&+\om_1\om_2^6\ N^n \del_n\om_2\ \p^2+\frac{2k_{10}^2}{\sqrt{h}}\  \call^{counter}[\p]
\biggr)\,,
\end{split}
\eqlabel{breg}
\end{equation}
where we used Stoke's theorem and, as prescribed in \cite{ss}, maintained 
only the boundary $\del\calm_6^r$ contribution, $N^n$ is a unit outward normal to $\del\calm_6^r$
and $h=\det(h_{ij})$ is a determinant of the induced metric on $\del\calm_6^r$.
The counter-term lagrangian $\call^{counter}$ should be constructed in such a way as to remove 
any divergent and $\w$-independent contributions from the effective boundary action in the limit 
$\del\calm_6^r\to\del\calm_6$. Given \eqref{solff}, it is easy to see that the latter is achieved with 
\begin{equation}
\call^{counter}[\p]=-\frac{1}{{2k_{10}^2}}\sqrt{h}\om_1\om_2^6\ N^n \del_n\om_2\ \p^2\,.
\eqlabel{lcou}
\end{equation} 
It is a very 
interesting  problem  to represent  an appropriate counter-term as a local functional 
of boundary metric and matter fields invariants. Though many impressive results in this direction
are obtained for large classes of specific supergravity backgrounds (see \cite{sk} and references therein), 
local counterterm expressions for  supergravity backgrounds as generic as we are discussing here are
not known. Clearly, the strong form of the gauge theory/string theory correspondence 
(in the context of four-dimensional renormalizable gauge theories with or without 
supersymmetry) implies that such local representation must exist.  As we demonstrate shortly, 
counter-term lagrangian \eqref{lcou} leads to  finite correlation functions of the renormalized 
boundary action.  Thus, even though 
we don't know the divergent structure of the regularized boundary action \eqref{breg},
\eqref{lcou} must remove all present divergences of the latter.       
Substituting \eqref{solff} into  \eqref{breg} we can obtain $\calf^r(\w,y)$ 
\begin{equation}
\begin{split}
\calf^r(\w,y)=&-\frac{i\w Q}{4 k_{10}^2}\left(1+\calo\left(\frac{\w}{T}\right)\right)
\int_{\del\calm_6^r}d^5\xi\sqrt{h}\ \t^{-1}\om_1\om_2^7\ N^n\del_n\t\\
=&-\frac{i\w Q}{4 k_{10}^2}\left(1+\calo\left(\frac{\w}{T}\right)\right)
\int_{\del\calm_6^r}d^5\xi\sqrt{h}\ \om_2^8\ N^n\del_n\t \,,
\end{split}
\eqlabel{regk}
\end{equation}
where we recalled the definition of $\t$.
Eq.~\eqref{rel1} implies that
\begin{equation}
0=\int_{\calm_6^r}d^6y \sqrt{\tg}\ \nabla\left(\om_2^8\nabla\t\right)\,. 
\eqlabel{bulk}
\end{equation}
Application of Stoke's theorem to \eqref{bulk} leads to\footnote{The $n^m$ normal vector is pointing 
inward, hence the sign.}
\begin{equation}
\begin{split}
\int_{\del\calm_6^r}d^5\xi\sqrt{h}\ \om_2^8\ N^n\del_n\t=&\int_{\calh_5}d^5\xi\sqrt{h}\ \om_2^8\ n^n\del_n\t\\
=&\int_{\calh_5}d^5\xi\sqrt{h}\ \om_2^8\ (2\pi T)=2\pi T\ \frac{\cala_8}{V_3}\,,
\end{split}
\eqlabel{stt}
\end{equation}
where in the second line we used \eqref{rel3}, \eqref{area}. 
Thus, 
\begin{equation}
\begin{split}
\calf(\w,y)=&\lim_{\del\calm_6^r\to\del\calm_6} \calf^r(\w,y)=-\frac{i\w Q (2\pi T)\cala_8}{4 V_3k_{10}^2}\\
=&-\frac{i\w \cala_8}{4 V_3k_{10}^2}=-\frac{i\w s}{8\pi}\,,
\end{split}
\eqlabel{kenfin}
\end{equation}
where we used \eqref{defq}, \eqref{entdef}. From \eqref{Gr} we obtain quoted result \eqref{result}.

\section*{Acknowledgments}
I would like to thank 
European Centre for Theoretical Studies in Nuclear Physics and Related Areas
for hospitality during the initial stages of this work.
Research at  Perimeter Institute is supported in part 
by funds from NSERC of Canada. Further supported by an 
NSERC Discovery grant is gratefully acknowledged.

\section*{Appendix}
Here we discuss  effective action for the  perturbation of metric \eqref{metric}
\begin{equation}
ds_{10}^2\to ds_{10}^2+\delta g_{12}(t,y)\ dx^1dx^2\equiv  ds_{10}^2+2\p(t,y)\om_2^2(y)\ dx^1dx^2\,,
\eqlabel{metpert}
\end{equation}
in a class of  supergravity backgrounds specified in section 2. 
When applicable, we use notations and results of \cite{ba}.

Type IIB supergravity action reads 
\begin{equation}
\begin{split}
S_{IIB}=&\frac{1}{2k_{10}^2}\int d^{10}\xi\ \sqrt{-\hat{g}}
\biggl\{R_{10}-\frac{\del_M\tau\del^M\bar{\tau}}{2(\Im\tau)^2}
-\frac{G\cdot \overline{C}}{12}
-\frac{F^2_{(5)}}{4\cdot 5!}
\biggr\}\\
&+\frac{1}{8i k_{10}^2}\int\ C_{(4)}\wedge G
\wedge \overline{G}\,,
\end{split}
\eqlabel{iibfull}
\end{equation}
where  $C_{(4)}=4\w\ dt\wedge dx^1\wedge dx^2\wedge dx^3$, $F_{(5)}=4\calf_5$, also 
\begin{equation}
\begin{split}
&\tau=i\frac{1+\calb}{1-\calb}\,,\qquad f=\frac{1}{(1-\calb\calb^\star)^{1/2}}\,,\\
&G=f(1-\calb) G_3\,.
\end{split}
\eqlabel{moreredef}
\end{equation}
Above redefinition is convenient to utilize results of \cite{ba}.

We would like to evaluate  \eqref{iibfull} in the deformed metric \eqref{metpert} to quadratic order in $\p$.
We find
\begin{equation}
\sqrt{-\hat{g}}\to \sqrt{-\hat{g}}\left(1-\frac 12\p^2\right)\,,
\eqlabel{detper}
\end{equation} 
\begin{equation}
R_{10}\to R_{10}+R_{10}^{(\p^2)}\,,
\eqlabel{ri1}
\end{equation}
where 
\begin{equation}
\begin{split}
R_{10}^{(\p^2)}=&\om_1^{-2}\left\{\ft 12 \left(\del_t \p\right)^2-\del^2_t\left(\p^2\right)\right\}\\
&+\om_2^{-2}\left\{-\ft 12 \left(\nabla\p\right)^2+\nabla^2\left(\p^2\right)+\nabla(\ln\om_1)
\nabla\left(\p^2\right)+8\nabla(\ln\om_2)
\nabla\left(\p^2\right)\right\}\,.
\end{split}
\eqlabel{ri2}
\end{equation}
Since the axiodilaton vary only over $\calm_6$, its bulk action contribution is not affected
by the perturbation \eqref{metpert}
\begin{equation}
-\frac{\del_M\tau\del^M\bar{\tau}}{2(\Im\tau)^2}\to -\frac{\del_M\tau\del^M\bar{\tau}}{2(\Im\tau)^2}=-T^{(1)M}_M\,,
\eqlabel{axid}
\end{equation} 
where $T^{(1)}_{MN}$ is the energy momentum tensor of the axiodilaton (see eq.~(3.10) of \cite{ba}). 
Similarly we have 
\begin{equation}
-\frac{1}{12} G\cdot \overline{G}\to -\frac{1}{12} G\cdot \overline{G}=-2T^{(3)M}_M\,,
\eqlabel{3form}
\end{equation}
where $T^{(3)}_{MN}$ is the energy momentum tensor of three index antisymmetric tensor field 
(see eq.~(3.11) of \cite{ba}). One has to be careful with evaluation of the action of the self-dual 
5-form. A correct prescription to do this was explained in \cite{dg}. Thus \cite{ba},
\begin{equation}
-\frac{F^2_{(5)}}{4\cdot 5!}\to -\frac{F^2_{(5)}}{4\cdot 5!}\left(1+\p^2\right)
=8\om_1^{-2}\om_2^{-8}\ \left(\nabla\w\right)^2\left(1+\p^2\right)\,,
\eqlabel{5formp}
\end{equation}
also \cite{ba}
\begin{equation}
\begin{split}
&\frac{1}{8i k_{10}^2}\int\ C_{(4)}\wedge G\wedge \overline{G}
\to \frac{1}{8i k_{10}^2}\int\ C_{(4)}\wedge G\wedge \overline{G}\\
&=\frac{1}{8i k_{10}^2}\int\ 8\w\ dt\wedge dx^1\wedge dx^2\wedge dx^3\wedge
G\wedge \bar{G}\\
&=\frac{1}{2k_{10}^2}\int\ dtd^3x
\int_{\calm_6}d^6y\sqrt{\tg}\  \biggl(
\frac{i\w}{3} G\cdot \star_6\bar{G}
\biggr)\,.
\end{split}
I.~Papadimitriou and K.~Skenderis,
``Correlation functions in holographic RG flows,''
JHEP {\bf 0410}, 075 (2004)
[arXiv:hep-th/0407071].
\eqlabel{cs}
\end{equation}
Collecting \eqref{detper}-\eqref{cs} we find
\begin{equation}
\begin{split}
S_{IIB}\to S_{IIB}+S_{bulk}[\p]\,,
\end{split}
\eqlabel{bulk1}
\end{equation}
where 
\begin{equation}
\begin{split}
S_{bulk}[\p]=\int d^{10}\xi \om_1\om_2^9\sqrt{\tg}\left(R_{10}^{(\p^2)}-\frac 12\p^2
\left[R_{10}-T^{(1)M}_M-2T^{(3)M}_M+\frac{F^2_{(5)}}{4\cdot 5!}\right]\right)\,.
\end{split}
\eqlabel{bulk2}
\end{equation}
The trace of Einstein equations implies (the self-dual 5-form does not contribute)
\begin{equation}
R_{10}=T^{(1)M}_M+T^{(3)M}_M\,.
\eqlabel{traceen}
\end{equation}
Additionally, the $tt$-component of Einstein equations is \cite{ba}
\begin{equation}
\begin{split}
R_{tt}=&\frac 12\ \om_1^2 T^{(3)M}_M+4\om_2^{-8}\left(\nabla\w\right)^2\\
=&\frac 12\ \om_1^{2}\left(T^{(3)M}_M-\frac{F^2_{(5)}}{4\cdot 5!}\right)\,.
\end{split}
\eqlabel{ttcom}
\end{equation}
So we can rewrite \eqref{bulk2} as
\begin{equation}
\begin{split}
S_{bulk}[\p]=\int d^{10}\xi \om_1\om_2^9\sqrt{\tg}\left(R_{10}^{(\p^2)}+\p^2\om_1^{-2}R_{tt}
\right)\,.
\end{split}
\eqlabel{bulk3}
\end{equation}
Since 
\begin{equation}
R_{tt}=\om_1\om_2^{-9}\ \nabla\left(\om_2^7\nabla\om_1\right)\,,
\eqlabel{rictt}
\end{equation}
we identify \eqref{bulk3} with \eqref{ac1}.

\end{document}